\documentclass[pra,aps,twocolumn,groupedaddress,showpacs,floatfix]{revtex4}
\usepackage{amsmath,amssymb,multirow,epsfig,bm}

\newcommand{\beq}{\begin{equation}}
\newcommand{\eeq}{\end{equation}}
\newcommand{\bea}{\begin{eqnarray}}
\newcommand{\eea}{\end{eqnarray}}

\begin{document}
\title{Critical exponents of the semimetal-insulator transition in graphene: A Monte Carlo study}

\author{Joaqu\'{\i}n E. Drut$^1$ and Timo A. L\"ahde$^2$ }
\affiliation{$^1$Department of Physics, The Ohio State University, Columbus, OH 43210--1117, USA}
\affiliation{$^2$Department of Physics, University of Washington, Seattle, WA 98195--1560, USA}


\begin{abstract}
The low-energy theory of graphene exhibits spontaneous chiral symmetry breaking due to pairing of quasiparticles and 
holes, corresponding to a semimetal-insulator transition at strong Coulomb coupling. We report a Lattice 
Monte Carlo study of the critical exponents of this transition as a function of the number of Dirac flavors 
$N_f^{}$, finding $\delta = 1.25 \pm 0.05$ for $N_f^{} = 0$, $\delta = 2.26 \pm 0.06$ for $N_f^{} = 2$ and 
$\delta = 2.62 \pm 0.11$ for $N_f^{} = 4$, with $\gamma \simeq 1$ throughout. We compare our results with recent 
analytical work for graphene and closely related systems, and discuss scenarios for the fate of the chiral transition 
at finite temperature and carrier density, an issue of relevance for upcoming experiments with suspended graphene samples.
\end{abstract}

\date{\today}

\maketitle


Graphene, a single layer of carbon atoms arranged in a honeycomb lattice~\cite{GeimNovoselov,CastroNetoetal}, provides a 
building block for more complex allotropes such as graphite~(graphene sheets attached by van der Waals forces), 
fullerenes~(graphene spheres with pentagonal dislocations) and nanotubes~(cylindrically rolled-up graphene). In the absence 
of electron-electron interactions, the valence and conduction bands of graphene are connected by two inequivalent ``Dirac 
points'', around which the low-energy excitations are massless quasiparticles with a linear dispersion relation and a Fermi velocity of 
$v_F^{} \simeq c/300$~\cite{Lanzara1,Lanzara3}. Such a semimetallic band structure is, unfortunately, unsuitable for many 
electronic applications, which depend crucially on the ability to externally modify the conduction properties, as routinely 
done with semiconducting devices. The quest to engineer a bandgap in graphene has thus been propelled to the forefront of 
current research. Hitherto suggested solutions include gap formation due to interaction with a 
substrate~\cite{Lanzara3}, induction of strain~\cite{strain}, and geometric confinement by means of nanoribbons or 
quantum dots~\cite{nanoribbons}.

\begin{figure}[b]
\epsfig{file=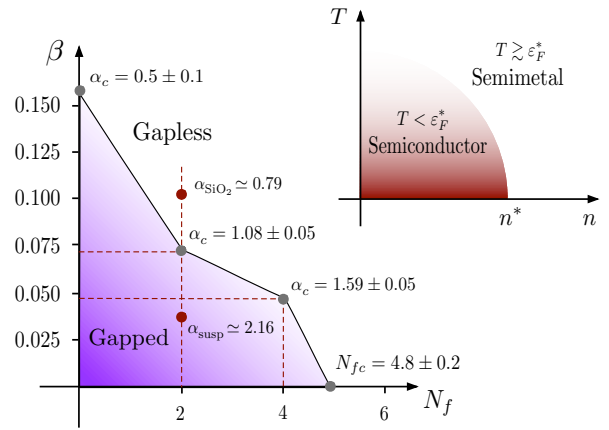, width=.9\columnwidth}
\caption{\label{fig:ph}(Color online) 
Phase diagram in the ($N_f^{}$ and $\beta$) plane. The gapped phase is bounded by a critical Coulomb coupling $\alpha_c^{} 
\equiv (4 \pi \beta_c)^{-1}$ and a critical number of fermion flavors $N_{fc}^{}$. The coupling on a $\mathrm{SiO_2}$ 
substrate is denoted by $\alpha^{}_\mathrm{SiO_2}$, and for suspended graphene by $\alpha^{}_\mathrm{susp}$. Inset: 
hypothetical phase diagram in the ($n$, $T$) plane. At low $T$, suspended graphene exhibits semimetallic properties whenever 
the carrier density $n$ exceeds a characteristic value $n^*_{}$. At the neutral point the semiconducting behavior persists 
until $T \simeq \varepsilon^*_F = \hbar v_F^{} \sqrt{\pi n^*}$, where the transition may be of 
Berezinskii-Kosterlitz-Thouless type or a crossover.}
\end{figure}

The low $v_F^{}$ in graphene indicates that the analog of the fine-structure constant of Quantum 
Electrodynamics~(QED) is $\alpha_g^{} \sim 1$, and thus the Coulomb attraction between electrons and holes may play a 
significant role in defining the ground-state properties. An intriguing possibility is that spontaneous formation of 
excitons (electron-hole bound states) and the concomitant breaking of chiral symmetry may turn graphene into a Mott 
insulator. While the strength of the Coulomb interaction precludes a perturbative approach, previous (approximate) analytic 
studies~\cite{Leal:2003sg} at the neutral point (zero carrier density~$n$) and zero temperature~$T$ have addressed the 
appearance of an excitonic gap as a function of $\alpha_g^{}$ (see Fig.~\ref{fig:ph}). Such treatments suggest that the 
transition into the insulating phase should be governed by essential singularities rather than power laws, a behavior known 
as Miransky scaling~\cite{Miranskyetal}.

In our recent Lattice Monte Carlo~(LMC) study~\cite{DrutLahde1}, indications were found that the chiral transition is of 
second order, with well-defined critical exponents. Subsequently, in Ref.~\cite{DrutLahde2} we provided a rough estimate of 
the critical exponents as $\delta \sim 2.3$, $\beta_m^{} \sim 0.8$ and $\gamma \simeq 1$, although a more precise 
determination was not possible due to insufficient data on large enough lattices. Nevertheless, Miransky scaling and 
classical mean-field exponents were found to be disfavored.

The aim of the present work is to provide a more rigorous and comprehensive determination of the quantum critical properties 
for $N_f^{} = 0, 2$ and $4$ Dirac flavors, as well as to contrast these results with recent analytical and simulational work 
for graphene and related theories. We also briefly elaborate on the mechanisms that inhibit exciton formation at non-zero 
$T$ and $n$, and their connection to other systems.

The LMC studies of Refs.~\cite{DrutLahde1,DrutLahde2,HandsStrouthos} suggest that the low-energy theory of 
graphene is an appropriate starting point for a quantitative analysis. This is defined by the Euclidean action
\begin{equation}
S_E^{} \,=\, - \int d^2x\,dt \: \bar\psi_a^{} D[A_0^{}] \psi_a^{}
+\,\frac{\varepsilon_0^{}}{2e^2} \int d^3x\,dt \: (\partial_i^{} A_0^{})^2,
\label{SE}
\end{equation}
with the Dirac operator
\begin{eqnarray}
D[A_0^{}] &=& \gamma_0^{} (\partial_0^{} + iA_0^{}) + v\gamma_i^{} \partial_i^{} + m_0^{}\openone, 
\quad i=1,2 \quad
\label{Dop}
\end{eqnarray}
where the $\psi_a^{}$ with $a = 1, \ldots, N_f^{}$ are four-component spinors in 2+1~dimensions,
$A_0^{}$ is a Coulomb field in 3+1 dimensions, and the case of a graphene monolayer is recovered for $N_f^{}=2$ in the
limit $m_0^{} \to 0$. Furthermore, $\alpha_g^{} \equiv e^2/(4 \pi v \varepsilon_0^{})$ with the inverse coupling $\beta 
\equiv v \varepsilon_0^{}/e^2$, such that screening by a substrate is reflected in the dielectric constant 
$\varepsilon_0^{}$. 

The gauge term of Eq.~(\ref{SE}) is discretized in the non-compact formulation~\cite{DrutLahde1,DrutLahde2}. The staggered 
discretization~\cite{Kogut-Susskind} of the fermionic component of Eq.~(\ref{SE}) is preferred, as chiral symmetry is then 
partially retained at finite lattice spacing. As $N$ staggered flavors correspond to $N_f^{} = 2N$ continuum Dirac 
flavors~\cite{BurdenBurkitt}, the case of $N_f^{} = 2$ is recovered for $N = 1$, giving
\begin{eqnarray}
S^f_E[\bar{\chi},\chi,U_0^{}] &=&
-\sum_{\bm{m},\bm{n}}
\bar\chi_{\bm{m}}^{} \: K_{\bm{m},\bm{n}}^{}[U_0^{}] \: \chi_{\bm{n}}^{},
\label{Sf}
\end{eqnarray}
where the $\chi_{\bm{n}}^{}$ are staggered fermion spinors, and the site indices $(\bm{m},\bm{n})$ are restricted to a 
2+1~dimensional sublattice. Invariance under spatially uniform, time-dependent gauge transformations is retained by the link 
variables $U^{}_{0,\bf n} = U^{}_{\bf n} \equiv \exp(i\theta^{}_{\bf n})$, where $\theta^{}_{\bf n}$ is the lattice gauge 
field. For $v = 1$, the staggered form of Eq.~(\ref{Dop}) is
\begin{eqnarray}
K_{\bm{m},\bm{n}}^{}[U] &\!\!=\!\!&
\frac{1}{2} \left[\delta_{\bm{m}+\bm{e}_0^{},\bm{n}}^{}\,U_{\bm{m}}^{}
- \delta_{\bm{m}-\bm{e}_0^{},\bm{n}}^{}\,U_{\bm{n}}^{\dagger}\right]
\label{Df} \\
&\!\!+\!\!&\frac{1}{2} \sum_i \eta_{i,\bm{m}}^{} \left[\delta_{\bm{m}+\bm{e}_i^{},\bm{n}}^{}
- \delta_{\bm{m}-\bm{e}_i^{},\bm{n}}^{}\right]
+ m_0^{}\,\delta_{\bm{m},\bm{n}}^{}, \nonumber
\end{eqnarray}
where $\eta_{1,\bm{n}}^{} = (-1)^{n_0^{}}$ and $\eta_{2,\bm{n}}^{} = (-1)^{n_0^{} + n_1^{}}$. Our simulations use the Hybrid 
Monte Carlo~(HMC) algorithm with $N$ pseudofermion flavors on a 2+1~dimensional space-time lattice of extent $L$, such that 
$\theta$ also propagates in the third spatial dimension of extent $L_z^{}$. Further details are given in 
Refs.~\cite{DrutLahde2,Rothe}.

We now seek to characterize the critical exponents of the chiral transition in graphene. The spontaneous breakdown of chiral 
symmetry in Eq.~(\ref{SE}) is signaled by a non-zero condensate $\sigma \equiv \langle \bar\chi \chi \rangle$. The mass term 
in Eq.~(\ref{Dop}) breaks chiral symmetry explicitly, generating a non-vanishing condensate, which is otherwise not possible 
at finite volume. The appearance of a gap in the quasiparticle spectrum of graphene at a critical coupling $\beta_c^{}$ is 
then marked by $\sigma \neq 0$ for $m_0^{} \to 0$. However, the ``chiral limit" $m_0^{} \to 0$ cannot be approached 
directly, as that limit corresponds to a very large fermionic correlation length, especially in the vicinity of $\beta_c^{}$ 
due to the appearance of Goldstone modes. Practical simulations are performed at finite $m_0^{}$, such that the limit 
$m_0^{} \to 0$ is reached by extrapolation, for which it is useful to also study the susceptibility $\chi_l^{} \equiv 
\partial \sigma/\partial m_0^{}$ and the logarithmic derivative $R \equiv \partial \ln\sigma/\partial \ln m_0^{}$. An 
instructive way to determine $\beta_c^{}$ and the critical exponents is by fitting an equation of state~(EOS) $m_0^{} = 
f(\sigma,\beta)$ to simulation data at finite $m_0^{}$. Knowledge of $f(\sigma,\beta)$ with good precision close to the 
transition then allows for an educated extrapolation to the chiral limit.


\begin{figure}[b]
\epsfig{file=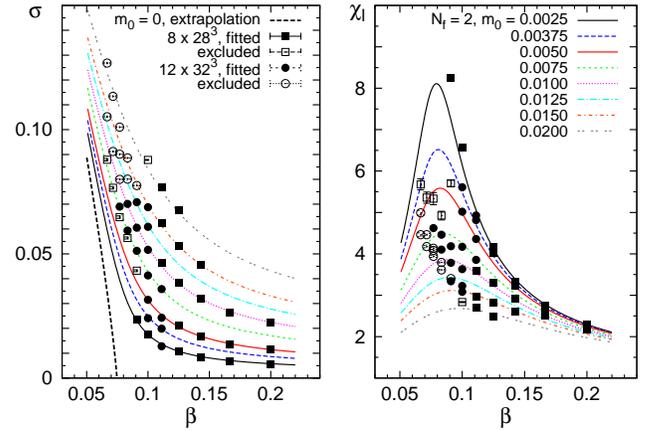, width=\columnwidth}
\caption{(Color online)
Chiral condensate $\sigma$ (left panel) and susceptibility $\chi_l^{}$ (right panel) for $N_f^{} = 2$. Data for $L = 28, L_z^{} = 8$ are indicated by squares, and for $L = 32, 
L_z^{} = 12$ by circles. The lines represent a $\chi^2_{}$ fit to $\sigma$ and $\chi_l^{}$ and extrapolation $m_0^{} \to 0$ 
using Eq.~(\ref{EOS}). The open datapoints are excluded due to finite-volume or lattice spacing effects. The optimal fit is 
$\beta_c^{} = 0.0738 \pm 0.0010$ and $\delta = 2.23 \pm 0.06$, with $X_0^{} = 0.36 \pm 0.05$, $X_1^{} = -0.13 \pm 
0.02$ and $Y_1^{} = -0.15 \pm 0.02$. The errors are of statistical origin. The method of analysis is described in detail in Refs.~\cite{DrutLahde1,DrutLahde2}.
\label{fig:cond_Nf2}}
\end{figure}

\begin{figure}[t]
\epsfig{file=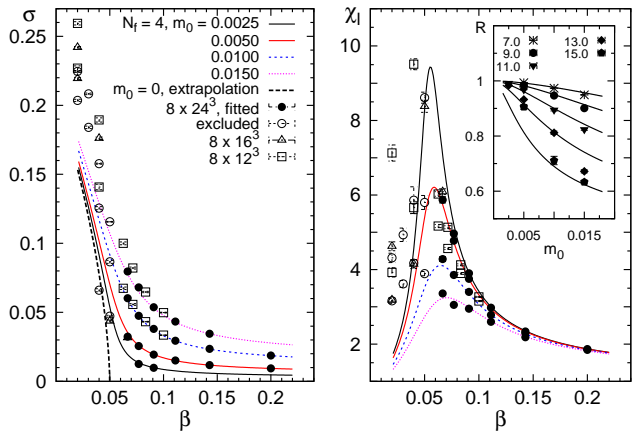, width=\columnwidth}
\caption{(Color online)
Chiral condensate $\sigma$ (left panel) and susceptibility $\chi_l^{}$ (right panel) for $N_f^{} = 4$, for 
lattices up to $L = 24, L_z^{} = 8$. Inset: logarithmic derivative $R$ for different $\beta^{-1}$. The optimal fit is 
$\beta_c^{} = 0.0499 \pm 0.0010$ and $\delta = 2.62 \pm 0.11$, with $X_0^{} = 0.19 \pm 0.05$, $X_1^{} = -0.09 \pm 0.02$ and 
$Y_1^{} = -0.08 \pm 0.02$. See also Fig.~\ref{fig:cond_Nf2}.
\label{fig:cond_Nf4}}
\end{figure}

\begin{figure}[t]
\epsfig{file=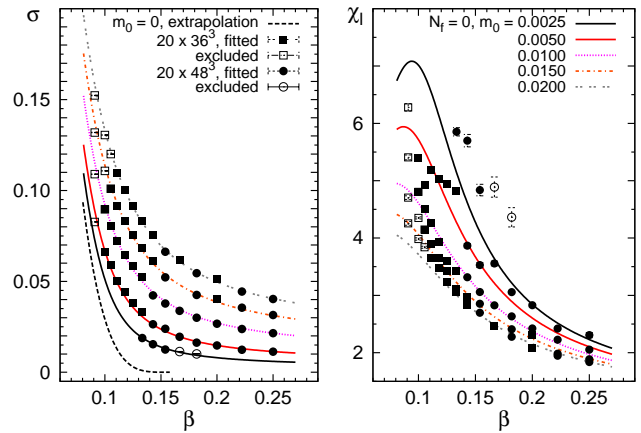, width=\columnwidth}
\caption{(Color online)
Chiral condensate $\sigma$ (left panel) and susceptibility $\chi_l^{}$ (right panel) for $N_f^{} = 0$, with $L = 36, 
L_z^{} = 20$ (squares) and $L = 48, L_z^{} = 20$ (circles). The optimal fit, with fixed $\delta = 1.25$ is $\beta_c^{} = 
0.158 \pm 0.001$, $X_0^{} = 0.16 \pm 0.02$, $X_1^{} = -0.10 \pm 0.05$ and $Y_1^{} = -0.11 \pm 0.02$. The range $\delta = 
1.25 \pm 0.05$ yields $\beta_c^{} = 0.16 \pm 0.02$.
\label{fig:cond_Nf0}}
\end{figure}

\begin{figure}[b]
\epsfig{file=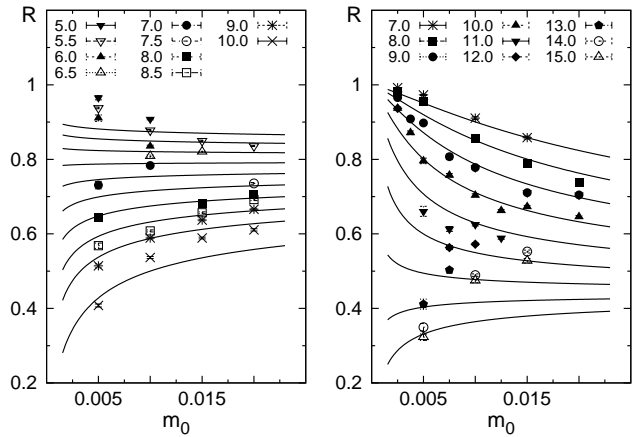, width=\columnwidth}
\caption{
Logarithmic derivative $R$ for $N_f^{} = 0$ (left panel) and $N_f^{} = 2$ (right panel), for different $\beta^{-1}$. The 
data for $N_f^{} = 0$ indicates that $\delta \simeq 1.2$, while the case of $N_f^{} = 2$ is compatible with $\delta \simeq 
2.2$. Finite-volume effects are substantial for $N_f^{} = 0$, at large $\beta$ and small $m_0^{}$.
\label{fig:R_Nf2}}
\end{figure}

We have considered the EOS successfully 
applied~\cite{Gockeleretal} to lattice QED,
\begin{eqnarray}
m_0^{}X(\beta) &=& Y(\beta) f_1^{}(\sigma) + f_3^{}(\sigma),
\label{EOS}
\end{eqnarray}
where $X(\beta)$ and $Y(\beta)$ are expanded around $\beta_c^{}$ such that
$X(\beta) = X_0^{} + X_1^{}(1-\beta/\beta_c^{})$ and $Y(\beta) = Y_1^{}(1-\beta/\beta_c^{})$.
The dependence on $\sigma$ is given by $f_1^{}(\sigma) = \sigma^b_{}$ and $f_3^{}(\sigma) = \sigma^\delta_{}$, where $b 
\equiv \delta - 1/\beta_m^{}$. The critical exponents are
\bea
\label{exponents_beta}
\beta_m^{} &\equiv& \left. \frac{\partial \ln \sigma}{\partial \ln (\beta_c^{} - \beta)} 
\right|_{m_0^{} = 0}^{\beta \nearrow \beta_c^{}},
\eea
and
\bea
\gamma \,\equiv -\left. \frac{\partial \ln \chi^{}_l}{\partial \ln (\beta_c^{} - \beta)} 
\right|_{m_0^{} = 0}^{\beta \nearrow \beta_c^{}}, \quad
\delta \,\equiv \left. \left [ \frac{\partial \ln \sigma}{\partial \ln m_0^{}} \right]^{-1}
\right|^{\beta = \beta_c^{}}_{m_0^{} \to 0},
\eea
which are assumed to obey the hyperscaling relation $\beta_m^{}(\delta - 1) = \gamma$. It should also be noted that $R \to 
1/\delta$ for $m_0^{} \to 0$ at $\beta = \beta_c^{}$. Our results and analyses in terms of Eq.~(\ref{EOS}) are 
shown in Fig.~\ref{fig:cond_Nf2} for $N_f^{} = 2$, in Fig.~\ref{fig:cond_Nf4} for $N_f^{} = 4$, and for the quenched case 
$N_f^{} = 0$ in Fig.~\ref{fig:cond_Nf0}. All of our results are consistent with $b = 1.00 \pm 0.05$, hence we conclude that 
$\gamma \simeq 1$ based on the hyperscaling relation, such that the remaining exponent to determine is $\delta$. Based on 
the EOS analysis and the logarithmic derivative $R$~(see Fig.~\ref{fig:R_Nf2}), we find $\delta = 2.26 \pm 0.06$ for $N_f^{} 
= 2$, $\delta = 2.62 \pm 0.11$ for $N_f^{} = 4$, and $\delta = 1.25 \pm 0.05$ for $N_f^{} = 0$. We observe that finite 
volume effects decrease with increasing $N_f^{}$ and that datapoints for small $\beta$ and large $m_0^{}$ in the broken 
phase are not well described by Eq.~(\ref{EOS}), likely due to a small correlation length associated with a growing 
excitonic gap.

An increase in $\delta$ with $N_f^{}$ is consistent with the LMC results of 
Ref.~\cite{HandsStrouthos} where a similar trend was found, culminating at $3.6 \lesssim \delta \lesssim 6$ for $N_f^{} = 
N_{fc}^{} \simeq 4.8$ where the chiral transition disappears. Such behavior is reminiscent of the Thirring model in 
$2+1$~dimensions~\cite{Christofietal}, where $\delta \simeq 2.8$ for $N_f^{} = 2$, reaching $\delta \simeq 7$ at a critical 
flavor number of $N_{fc}^{} \simeq 6.6$. Extensive LMC studies of QED have found $\delta \sim 2.2$ for 
$N_f^{} = 0$~\cite{QuenchedQED}, while for QED with dynamical fermions $\delta \simeq 3$~\cite{Gockeleretal}. 
The case of QED in $2+1$~dimensions (QED$_3^{}$) is noteworthy as the LMC study of Ref.~\cite{HandsQED3} yielded 
$\delta \simeq 2.3$ for $N_f^{} = 1$ and $\delta \simeq 2.7$ for $N_f^{} = 4$, which are suggestive of our values for 
graphene, although spontaneous chiral symmetry breaking in QED$_3^{}$ is difficult to establish as the order parameter can 
be exponentially suppressed for large $N_f^{}$.

The gap-equation analysis of Ref.~\cite{Leal:2003sg} reported $\beta_c^{} \simeq 0.16$ 
for $N_f^{} = 0$ and $\beta_c^{} \simeq 0.066$ for $N_f^{} = 2$, which are in qualitative agreement with our results. 
However, the transition of Ref.~\cite{Leal:2003sg} is of infinite order and vanishes for $N_f^{} = 4$. These discrepancies 
are smallest for $N_f^{} = 0$, where our results approach $\delta = 1$. Our observations are thus in line with 
indications~\cite{SDE} that the critical exponents, as obtained from Schwinger-Dyson equation (SDE) analyses, 
may be dependent on the chosen resummation scheme.

An effective theory containing both the order parameter and the Dirac quasiparticles as dynamical fields has recently been 
developed in Ref.~\cite{HJV}. Based on an expansion to leading order around $\epsilon=3-d$ spatial dimensions, the 
long-range $\sim 1/r$ Coulomb tail was found to be irrelevant in the renormalization group (RG) sense, such that the chiral 
transition could then be described using only short-range interactions of the Gross-Neveu-Yukawa form, yielding the 
estimates $\gamma \sim 1.25$ and $\delta \sim 2.8$ for the critical exponents at $N_f^{} = 2$, in qualitative agreement with 
our present findings, as well as with large-$N_f^{}$ calculations of the RG flow~\cite{FourFermi}. However, our results are
not compatible with $\delta = 2 + O(1/N_f^{})$ found in Refs.~\cite{HerbutPRL,HJV}, which is surprising as the 
Gross-Neveu-Yukawa theory is expected to interpolate between $\delta=19/5 \simeq 4$ at $N_f^{}=0$ and $\delta = 2$ in the 
$N_f^{} \to \infty$ limit \cite{HJV}. It is not clear, as no chiral transition exists in the graphene theory above the 
critical flavor number $N_{fc}^{} = 4.8$~\cite{HandsStrouthos}, how to consistently compare our results with 
large-$N_f^{}$ estimates.


What is the fate of the semimetal-insulator transition at non-zero temperature? On the basis of the Mermin-Wagner 
theorem~\cite{MerminWagner}, one expects either a crossover or a Berezinskii-Kosterlitz-Thouless~(BKT) 
transition~\cite{RMPSondhi} at a critical temperature $T_c^{}$. The most compelling experimental evidence so far for a 
BKT transition has been reported in Ref.~\cite{FiniteB_Exp}, where graphene samples on a substrate were subjected to 
transverse magnetic fields up to $B \simeq 30$~T. In the temperature range of $10$~K to $1$~K, a growth in the resistivity 
by a factor of $\sim 200$ was observed, and attributed to the ``magnetic catalysis" predicted in 
Refs.~\cite{FiniteB_Kh,FiniteB_Gorbar}. At $B = 0$, the resistivity of annealed suspended graphene was 
observed~\cite{Bolotin2} to increase by a factor of $\sim 3$ over a temperature range of $200$~K to $50$~K, while also 
changing character from metallic to semiconducting. A study of the low-energy theory of graphene at non-zero $T$ is thus 
clearly called for, possibly along the lines of Ref.~\cite{HandsGNM}, which considered the Gross-Neveu model in 
$2+1$~dimensions. 

At low $T$, the large extent of the imaginary time dimension renders the system effectively three-dimensional, such that a 
chiral transition may still be observed at a critical~density~$n^*$.
Interestingly, away from the neutral (unpolarized) point, non-relativistic Fermi systems (such as the asymmetric Fermi 
liquid in the context of ultracold atoms and dilute neutron matter~\cite{AFL}) can undergo transitions into exotic 
phases~\cite{FFLO, Sarma, BP} before reverting to a fully polarized normal state. Whether the low-energy theory of graphene 
exhibits such phenomena is currently unknown.


We acknowledge support under U.S. DOE Grants No.~DE-FG-02-97ER41014, No.~DE-FG02-00ER41132, and No.~DE-AC02-05CH11231, UNEDF 
SciDAC Collaboration Grant No.~DE-FC02-07ER41457 and NSF Grant No.~PHY--0653312. J. E. D. acknowledges the hospitality of 
the Institute for Nuclear Theory during the completion of this work. This work was supported in part by an allocation of 
computing time from the Ohio Supercomputer Center. We thank A.~Bulgac and M.~J.~Savage for computer time, and 
R.~J.~Furnstahl, S.~J.~Hands, I.~Herbut and D.~T.~Son for valuable comments.


\end{document}